\documentclass[a4paper,11pt]{article}
\usepackage{pos}

\title{$t\bar{t}jj$ - NLO QCD corrections to top quark pair production and decays at the LHC}

\author*[a]{Daniel Stremmer}

\affiliation[a]{ Institute for Theoretical Particle Physics
and Cosmology, RWTH Aachen University, \\D-52056 Aachen, Germany}

\emailAdd{daniel.stremmer@rwth-aachen.de}

\abstract{
In these proceedings we present the calculation of NLO QCD corrections to $pp\to t\bar{t}jj$ in the dilepton decay channel. The narrow width approximation is used to model the decays of the top quark pair preserving spin correlations. Jet radiation and QCD corrections are consistently included in the production and decay of the top quarks. We discuss the size of NLO QCD corrections and the main theoretical uncertainties of fiducial cross sections at the integrated and differential level. In addition, we examine the contributions of jet radiation in the production and decay of the top-quark pair, as well as of a mixed configuration where jet radiation is present simultaneously in the production and decay stages.

\bigskip

P3H-23-061, \, TTK-23-23
}

\FullConference{16th International Symposium on Radiative Corrections: Applications of Quantum Field Theory to Phenomenology (
RADCOR2023)\\
28th May - 2nd June, 2023\\
Crieff, Scotland, UK\\}

\begin{document}
\maketitle

\section{Introduction}
Although the Higgs production in association with a top-quark pair ($t\bar{t}H$) is only $1\%$ of the total Higgs cross section at the LHC, this process is of high relevance for the Higgs program at the LHC. Due to its structure, it is a direct probe of the top-quark Yukawa coupling ($Y_t$) already at the tree level. In 2018 the observation of $t\bar{t}H$ was reported by the ATLAS and CMS collaborations \cite{ATLAS:2018mme,CMS:2018uxb} and even recently single-channel observations in the $H\to \gamma\gamma$ decay channel were possible \cite{CMS:2020htp,ATLAS:2020ior} although its small branching ratio. In contrast, in the Higgs decay channel with the largest branching ratio, $H\to b\bar{b}$, such precise measurements are not yet possible due to the large irreducible background from the prompt production of a bottom-quark pair ($t\bar{t}b\bar{b}$) and the enormous reducible background from top-quark production with two additional light jets ($t\bar{t}jj$) leading to significant systematic uncertainties.
In addition, the study of the cross-section ratios $R_b=\sigma_{t\bar{t}b\bar{b}}/\sigma_{t\bar{t}jj}$ and $R_c=\sigma_{t\bar{t}c\bar{c}}/\sigma_{t\bar{t}jj}$ can be used for powerful tests of the efficiency of $b$/$c$ tagging algorithms in a complex environment with many jets from different production mechanisms. Such an measurement was already performed by the CMS collaboration and differences up to $2.5\sigma$ has been found between theoretical predictions and measurements of $R_b$ \cite{CMS:2020utv}.

While for $t\bar{t}b\bar{b}$ theoretical predictions with matrix elements accurate at NLO QCD in both the production and decay of the top-quark pair are available in the literature \cite{Denner:2020orv,Bevilacqua:2021cit,Bevilacqua:2022twl}, the situation is less advanced for $t\bar{t}jj$ where NLO QCD corrections are only known for stable top quarks \cite{Bevilacqua:2010ve,Bevilacqua:2011aa}.

In this proceeding we discuss the first steps towards a more realistic description of $t\bar{t}jj$ accurate at NLO QCD in both the production and decay of the top-quark pair. We consider the dilepton decay channel and perform the decays of the top quarks and $W$ gauge bosons in the narrow-width approximation (NWA), i.e. in the limit $\Gamma/m\to 0$. We consistently include NLO QCD corrections as well as jet radiation in both the production and decay of the top-quark pair. Furthermore, we discuss the effects of jet radiation in the production and decay of the top-quark pair, as well as of a mixed configuration where light radiation is present in both decay stages simultaneously.

\section{Setup of the calculation}

\begin{figure}[t!]
  \begin{center}
  \includegraphics[trim= 20 700 20 20, width=\textwidth]{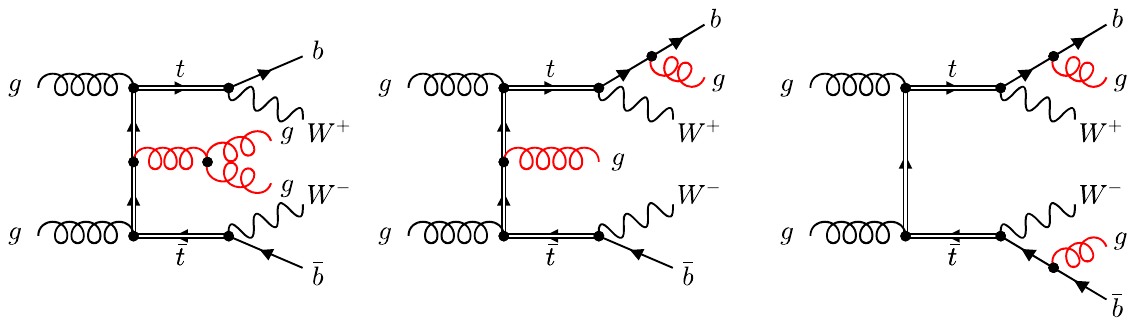}
\end{center}
  \caption{\label{fig:fd-LO} \it  Representative Feynman diagrams for the {\it Prod.}, {\it Mix} and {\it Decay} contributions at LO with suppressed $W$ gauge boson decays. Feynman diagrams were produced with the help of the \textsc{FeynGame} program \cite{Harlander:2020cyh}.}
\end{figure}

In this section we discuss the main points of the computational setup for the calculation of NLO QCD corrections to $t\bar{t}jj$ in the dilepton decay channel. The full setup can be found in Ref. \cite{Bevilacqua:2022ozv}. As already discussed in the introduction, the decays of the top quarks and $W$ gauge bosons are performed in the NWA leading to the following decay chain at LO at the order $\mathcal{O}(\alpha_s^4\alpha^4)$
\begin{equation}
pp \to t\bar{t}(jj)\to W^+W^-\,b\bar{b}jj \to \ell^+\nu_{\ell}\, \ell^-\bar{\nu}_{\ell} \, b\bar{b}\,jj +X, 
\end{equation}
with $\ell^{\pm}=\mu^{\pm},e^{\pm}$ and where the brackets indicate that the light jets can be emitted from both the production and decay of the top-quark pair. At this order the process can be uniquely divided into three resonant contributions based on the origin of the light jets according to 
\begin{equation}
\label{eq_LO}
\begin{split}
    d\sigma^{\textrm{LO}}_{t\bar{t}jj}&=\Gamma_{t}^{-2}\bigl(
    \overbrace{d\sigma_{t\bar{t}jj}^{\textrm{LO}}\,
    d\Gamma_{t\bar{t}}^{\textrm{LO}}}^{\rm Prod.}
    +\overbrace{d\sigma_{t\bar{t}}^{\textrm{LO}}
    \,d\Gamma_{t\bar{t}jj}^{\textrm{LO}} }^{\rm Decay}   
    +\overbrace{d\sigma_{t\bar{t}j}^{\textrm{LO}}
    \,d\Gamma_{t\bar{t}j}^{\textrm{LO}}}^{\rm Mix}
    \bigl).
    \end{split}
\end{equation}
In particular, we have the {\it Prod.} ({\it Decay}) contribution where light jets are emitted only in the production (decay) of the top-quark pair, and finally we have the {\it Mix} contribution where jet radiation is present in both decay stages. Example diagrams for the three resonant contributions are shown in Figure \ref{fig:fd-LO}.

In order to include NLO QCD corrections of the order $\mathcal{O}(\alpha_s^5\alpha^4)$, Eq. \eqref{eq_LO} has to be extended in the following way
\begin{equation}
\footnotesize
\label{eq_NLO}
\begin{split}
    d\sigma^{\textrm{NLO}}_{t\bar{t}jj}&=\Gamma_{t}^{-2}
    \bigl(
    \overbrace{
    \left(d\sigma_{t\bar{t}jj}^{\textrm{LO}}+ d\sigma_{t\bar{t}jj}^{\textrm{virt}}+
    d\sigma_{t\bar{t}jjj}^{\textrm{real}}\right)
    \,d\Gamma_{t\bar{t}}^{\textrm{LO}}}^{{\rm Prod.}}
     +\overbrace{d\sigma_{t\bar{t}}^{\textrm{LO}}
    \,\left(d\Gamma_{t\bar{t}jj}^{\textrm{LO}}
    +d\Gamma_{t\bar{t}jj}^{\textrm{virt}}+
    d\Gamma_{t\bar{t}jjj}^{\textrm{real}}\right)}^{{\rm Decay}}
    \\&
     +\underbrace{ 
     d\sigma_{t\bar{t}j}^{\textrm{LO}}
    \,d\Gamma_{t\bar{t}j}^{\textrm{LO}}
     +
     d\sigma_{t\bar{t}jj}^{\textrm{LO}}
    \,d\Gamma_{t\bar{t}}^{\textrm{virt}}+
    d\sigma_{t\bar{t}}^{\textrm{virt}}\,
    d\Gamma_{t\bar{t}jj}^{\textrm{LO}}+
    d\sigma_{t\bar{t}j}^{\textrm{virt}}\,
    d\Gamma_{t\bar{t}j}^{\textrm{LO}}+
    d\sigma_{t\bar{t}j}^{\textrm{LO}}\,
    d\Gamma_{t\bar{t}j}^{\textrm{virt}}+
     d\sigma_{t\bar{t}jj}^{\textrm{real}}\,
    d\Gamma_{t\bar{t}j}^{\textrm{real}}+
    d\sigma_{t\bar{t}j}^{\textrm{real}}\,
    d\Gamma_{t\bar{t}jj}^{\textrm{real}}}_{{\rm Mix}}
    \bigl),
    \end{split}
\end{equation}
where we directly have split the full calculation into the three finite resonant contributions. We note that this equation directly implies a mixing between the different resonant configurations at NLO QCD. The mixing is induced from the fact that QCD splittings in the production or decay of the top-quark pair from different Born resonant configurations can lead to the same set of real corrections. Such real corrections are included in the {\it Mix} contribution at NLO QCD where unresolved particles are allowed both in the production and decay of the top-quark pair but the total number of unresolved particles is still limited to one. In the next sections we will discuss this issue in more detail and present an alternative way to quantify the effects from jet radiation and NLO QCD corrections in top-quark decays.

The calculation is performed within the \textsc{Helac-Nlo} framework \cite{Bevilacqua:2011xh} consisting of the two programs \textsc{Helac-1Loop} \cite{Ossola:2006us,Ossola:2007ax,vanHameren:2009dr,Draggiotis:2009yb,vanHameren:2010cp} and \textsc{Helac-Dipoles} \cite{Czakon:2009ss}. The virtual corrections are cross-checked with the matrix element generator \textsc{Recola} \cite{Actis:2012qn,Actis:2016mpe,Denner:2016kdg} and for the real corrections we employ two independent subtraction schemes, the Catani-Seymour \cite{Catani:1996vz,Catani:2002hc} and Nagy-Soper scheme \cite{Bevilacqua:2013iha}, to ensure the correctness of our results. Both subtraction schemes have been extended to handle arbitrary QCD splittings inside decay processes.

We closely follow the event selection from a measurement on top-quark pair production with additional light jets by the CMS collaboration \cite{CMS:2022uae}. The full event selection and all input parameters can be found in Ref. \cite{Bevilacqua:2022ozv}. A key point of this setup is the choice of the cut $\Delta R_{jb} > 0.8$ which is used to suppress light jet radiation from the top-quark decays. Because of this, we additionally considered a second inclusive setup with a reduced cut of $\Delta R_{jb} > 0.4$ to investigate the size of the three resonant contributions in more detail. We employ the NNPDF3.1 NLO PDF set \cite{NNPDF:2017mvq} via the LHAPDF interface \cite{Buckley:2014ana} at LO and NLO QCD. The renormalisation ($\mu_R$) and factorisation ($\mu_F$) scales are set to a common scale ($\mu_0$) given by
\begin{equation}
\label{eq_scale}
\mu_R=\mu_F=\mu_0=\frac{H_T}{2}=\frac{1}{2}\left( \sum_{i=1}^2 p_{T {\ell_i}} + \sum_{i=1}^2 
        p_{T {j_i}} + \sum_{i=1}^2 p_{T {b_i}}+ p_T^{miss}\right),
\end{equation}
and scale uncertainties are obtained from a $7$-point scale variation around the central value.

\section{Integrated fiducial cross sections}
\begin{table}
  \centering
  \begin{tabular}{lllcc}
  \hline
     $i$     & $\sigma^{\rm LO}$ [fb]       &
     $\sigma^{\rm NLO}$ 
     [fb] &    $\sigma^{\rm LO}_i/\sigma^{\rm LO}_{\rm Full}$ 
     & $\sigma^{\rm NLO}_i/\sigma^{\rm NLO}_{\rm Full}$ \\
 &&&&\\
\hline
Full & $868.8(2)^{\, +60\%}_{\, -35\%}$ 
& $~1225(1)^{\, ~+1\%}_{\, -14\%}$ 
& $1.00$ & $~~1.00$ \\
Prod. & $ 843.2(2)^{\, +60\%}_{\, -35\%} $ & $ ~1462(1)^{\,+12\%}_{\,-19\%}$
& $0.97$ &  $~~1.19$ \\
Mix & $25.465(5)$ & $-236(1)$ & $0.029$ & $-0.19$\\
Decay & $0.2099(1)$ & $0.1840(8)$ & $0.0002$& $~~0.0002$\\
  \hline
  \end{tabular}
  \caption{\it \label{tab:1a} Integrated fiducial cross section at LO and NLO QCD for the $pp \to t\bar{t}jj$ process with $\Delta R_{jb} > 0.8$. Results are shown for the full calculation and for the three resonant contributions {\it Prod.}, {\it Decay} and {\it Mix}. Table was taken from \cite{Bevilacqua:2022ozv}.}
\end{table}
In Table \ref{tab:1a} we show the integrated fiducial cross section at LO and NLO QCD for the full calculation and divided into the three resonant regions for $\Delta R_{jb} > 0.8$. Scale uncertainties as well as statistical uncertainties from the phase space integration are also displayed. We find at LO that the full calculation is purely dominated by the {\it Prod.} contribution with $97\%$ of the full calculation and therefore the {\it Mix} and {\it Decay} configurations can be safely neglected compared to the scale uncertainties of $60\%$ at this order. At LO the $gg$ production channel is the largest one with about $65\%$ of the full calculation followed by the $gq$ channel with $31\%$ and the purely quark induced channel $qq'$ is the smallest one with only $4\%$. For the full calculation we find NLO QCD corrections at level of $40\%$ which are well within the LO scale uncertainties. In addition, the theoretical uncertainties obtained from scale variation are reduced by a factor of $4$ to $14\%$. The {\it Mix} contribution becomes more relevant at NLO QCD due to the mixing of the different resonant contributions. In particular, its sign changes and its relative size increases in absolute value from $3\%$ to $19\%$. However, the large negative contribution of {\it Mix} is mainly induced from the NLO QCD corrections to the top-quark decays of the {\it Prod.} configuration at LO. At last we have recalculated the integrated fiducial cross section with the MSHT20 \cite{Bailey:2020ooq} and CT18 \cite{Hou:2019efy} PDF sets. We find differences in the central value between these two PDF sets and our default one, NNPDF3.1, of about $1\%-3\%$. These differences are of the same size as the internal PDF uncertainties of the three PDF sets.

\begin{table}
  \centering
  \begin{tabular}{lllcc}
    \hline
     $i$     & $\sigma^{\rm LO}$ [fb]       &
     $\sigma^{\rm NLO}$ 
     [fb] &    $\sigma^{\rm LO}_i/\sigma^{\rm LO}_{\rm Full}$ 
     & $\sigma^{\rm NLO}_i/\sigma^{\rm NLO}_{\rm Full}$ \\
 &&&&\\
\hline
Full & $1074.5(3)^{\, +60\%}_{\, -35\%}$ 
& $~1460(1)^{\, ~+1\%}_{\, -13\%}$ 
& $1.00$ & $~~1.00$ \\
Prod. & $ 983.1(3)^{\, +60\%}_{\, -35\%} $ & $ ~1662(1)^{\,+11\%}_{\,-18\%}$
& $0.91$ &  $~~1.14$ \\
Mix & $89.42(3)$ & $-205(1)$ & $0.083$ & $-0.14$\\
Decay & $1.909(1)$ & $2.436(6)$ & $0.002$& $~~0.002$\\
   \hline
  \end{tabular}
  \caption{\it \label{tab:1b}  Same as in Table \ref{tab:1a} but for $\Delta R_{jb} > 0.4$. Table was taken from \cite{Bevilacqua:2022ozv}.}
\end{table}

As already motivated in the last section, we have performed the same calculation for a second setup by reducing the cut $\Delta R_{jb} > 0.8$ to $\Delta R_{jb} > 0.4$. The integrated fiducial cross section for this setup is shown in Table \ref{tab:1b}. First, we notice only a small dependence on this cut for {\it Prod.} since this contribution increases only slightly by $16\%$. However, for {\it Mix} and {\it Decay} the dependence is much stronger as these contributions increase by $250\%$ and $810\%$, respectively. Still, both contributions amount to only $8.3\%$ and $0.2\%$ of the full calculation at LO and thus are significantly smaller than the scale uncertainties. Due to the mixing at NLO QCD we find that the relative size of the {\it Mix} contribution is reduced in absolute value to $14\%$ from $19\%$ in the default setup.

An alternative way to quantify the effects of NLO QCD corrections and jet radiation in top-quark decays is the comparison of the full calculation with the {\it Prod.} contribution with LO top quark decays, which we call {\it Prod. LOdecay} and simply amounts to a rescaling of the {\it Prod.} result to the LO top-quark width. We find for our default setup with $\Delta R_{jb} > 0.8$ that this approximation leads to the same result as the full calculation in the central value of the integrated fiducial cross section. For the second setup with $\Delta R_{jb} > 0.4$ this approximation underestimates the full calculation by $5\%$. In addition, the scale uncertainties of the full calculation is reduced at the level of $5\%$ compared to this approximation for both setups.

\section{Differential fiducial cross sections}
\begin{figure}[t!]
  \begin{center}
     \includegraphics[width=0.45\textwidth]{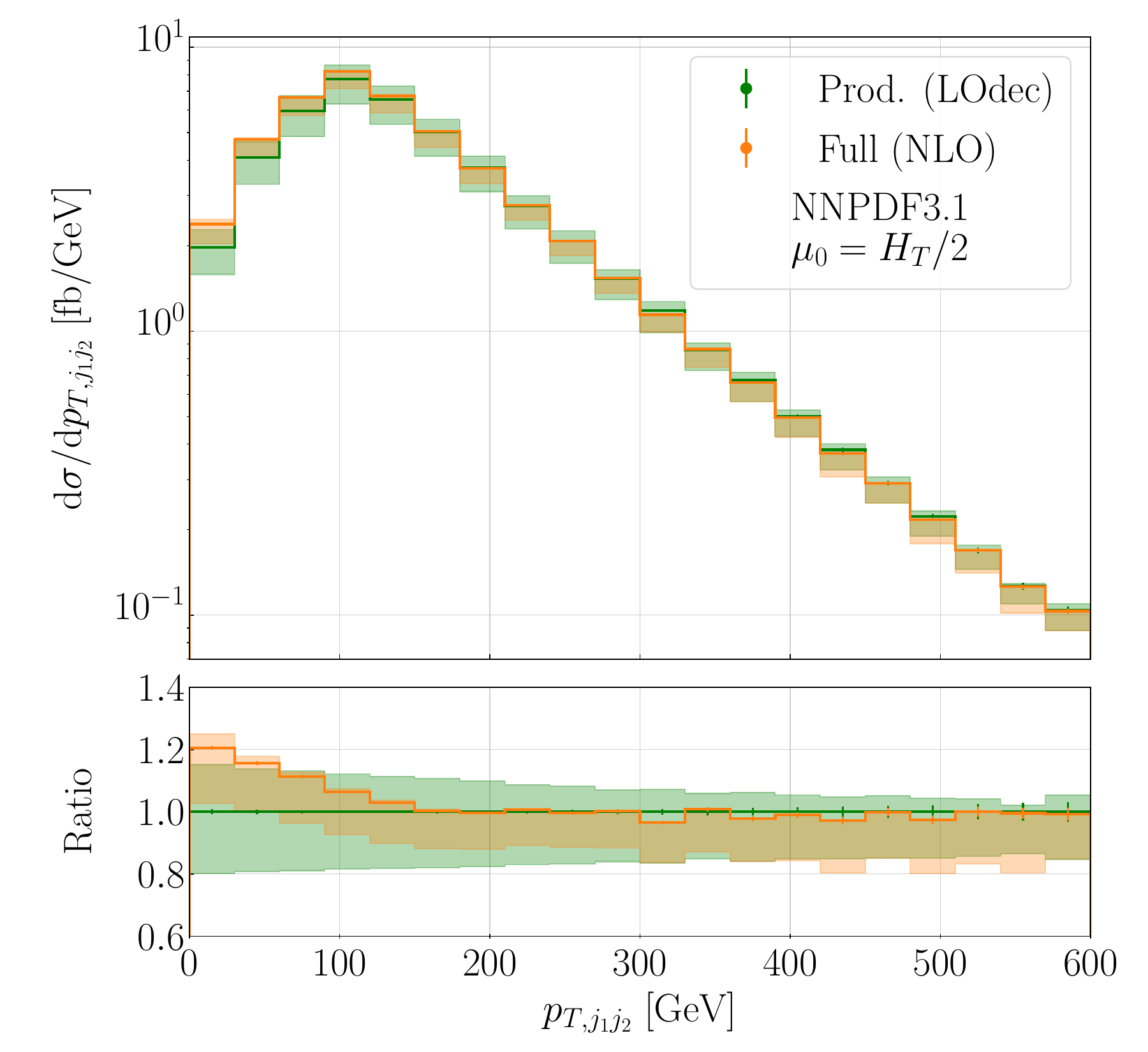}
      \includegraphics[width=0.45\textwidth]{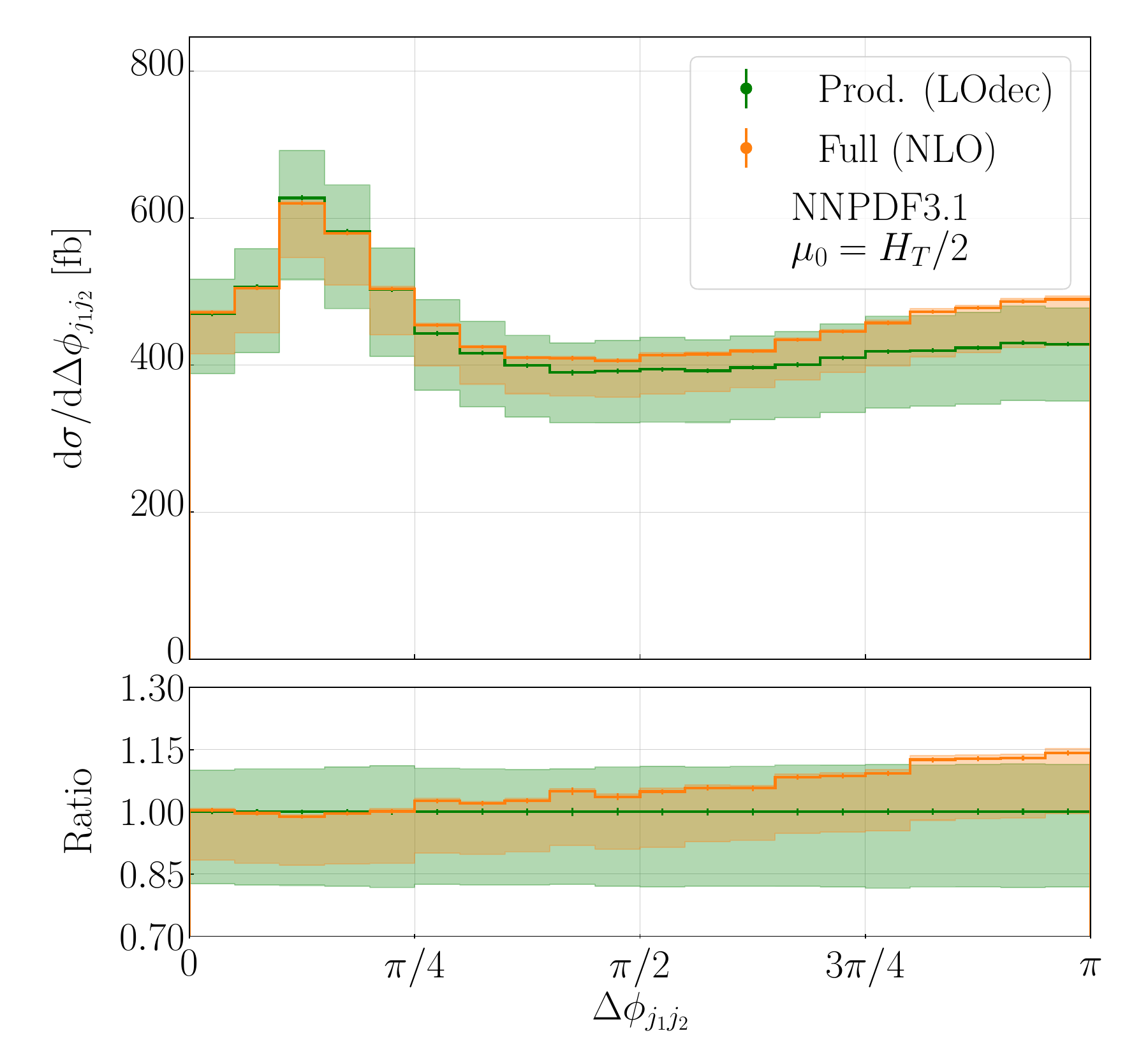}
\end{center}
\caption{\label{fig:LOdec} 
\it Differential cross-section distributions for the observables $p_{T, \, j_1j_2}$ and $\Delta \phi_{j_1j_2}$ employing the full calculation (orange) and the {\it Prod.} contribution with LO top-quark decays (green) with $\Delta R_{jb} >0.4$. Figures were taken from \cite{Bevilacqua:2022ozv}.}
\end{figure}

In the next step we have performed a comparison between the full calculation (orange) and the {\it Prod. LOdecay} approximation (green) at NLO QCD also at the differential level shown in Figure \ref{fig:LOdec} for the two observables $p_{T, \, j_1j_2}$ and $\Delta \phi_{j_1j_2}$ with $\Delta R_{jb} >0.4$. For the first observable $p_{T, \, j_1j_2}$ we find shape distortions between the two predictions by up to $20\%$ in the beginning of the spectrum which are reduced towards the tail where the two results become identical. Similar to the integrated level discussed in the previous section, the scale dependence in the full calculation is reduced for $p_{T, \, j_1j_2}<300$ GeV by $5\%$. Also for angular distribution like $\Delta \phi_{j_1j_2}$ shape distortions up to $15\%$ for large azimuthal angle differences are possible between the full calculation and the {\it Prod. LOdecay} approximation and again the scale dependence is reduced in the full calculation by $5\%$.

\begin{figure}[t!]
  \begin{center}
   \includegraphics[width=0.45\textwidth]{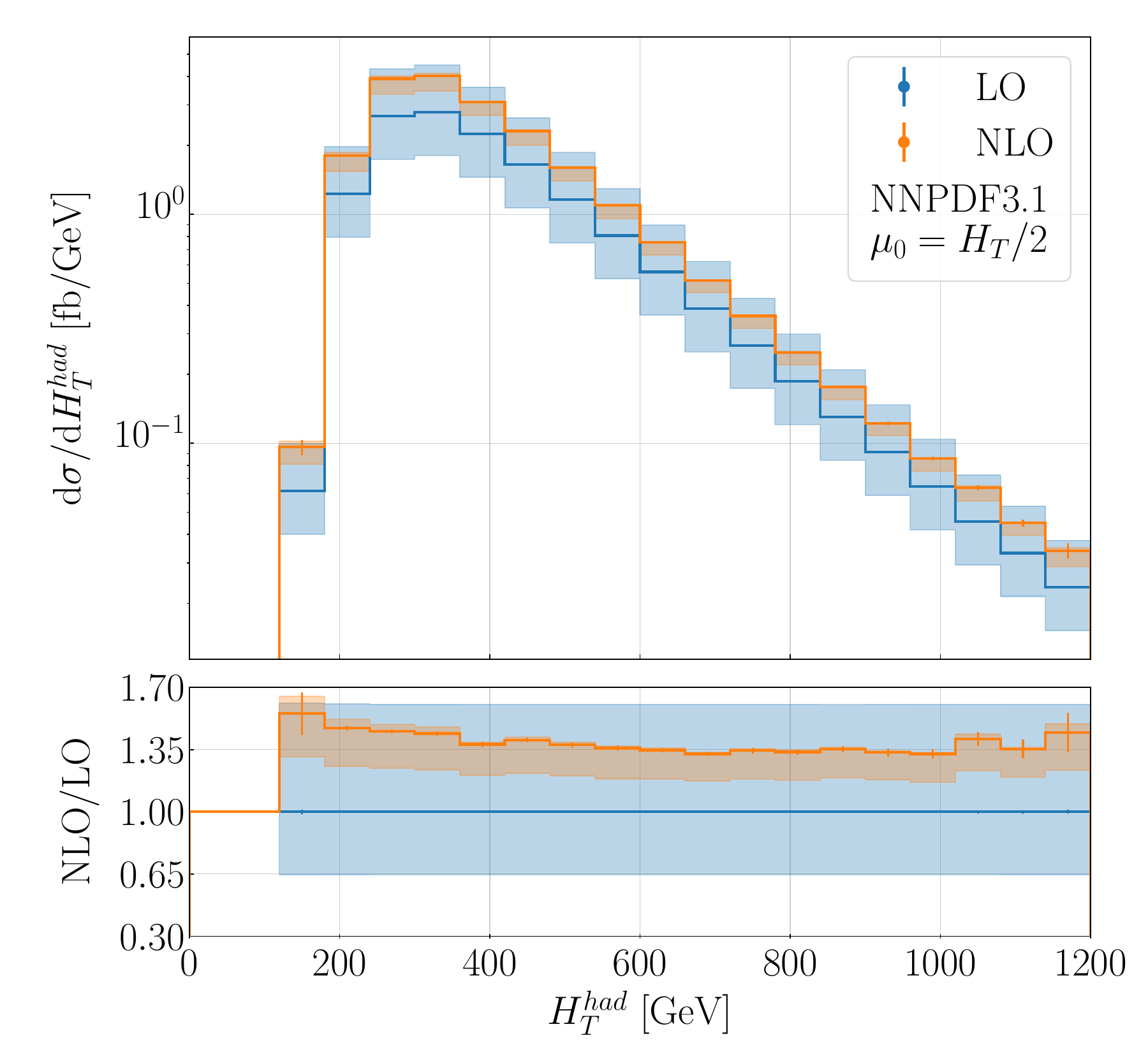}
   \includegraphics[width=0.45\textwidth]{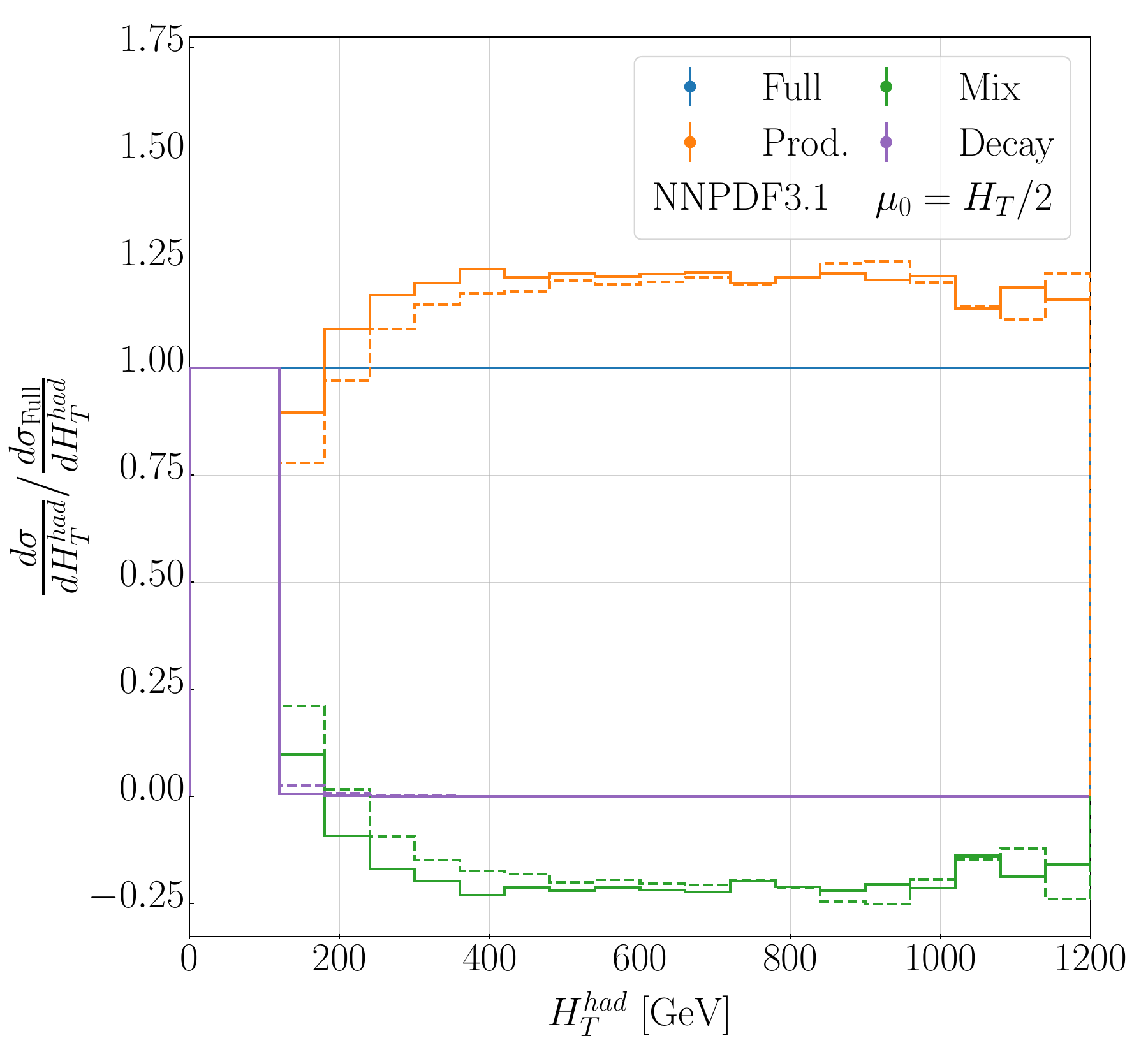}
\end{center}
\caption{\label{fig:kfac1} 
\it Left: Differential cross-section distribution for the observable $H_T^{had}$ at LO and NLO QCD with $\Delta R_{jb} >0.8$.
Right: Relative size of the three resonant contributions, {\it Prod.}, {\it Mix} and {\it Decay}, at NLO QCD for $\Delta R_{jb} >0.8$  (solid line) and  $\Delta R_{jb} >0.4$ (dashed line). Figures were taken from \cite{Bevilacqua:2022ozv}.}
\end{figure}

\begin{figure}[t!]
  \begin{center}
   \includegraphics[width=0.45\textwidth]{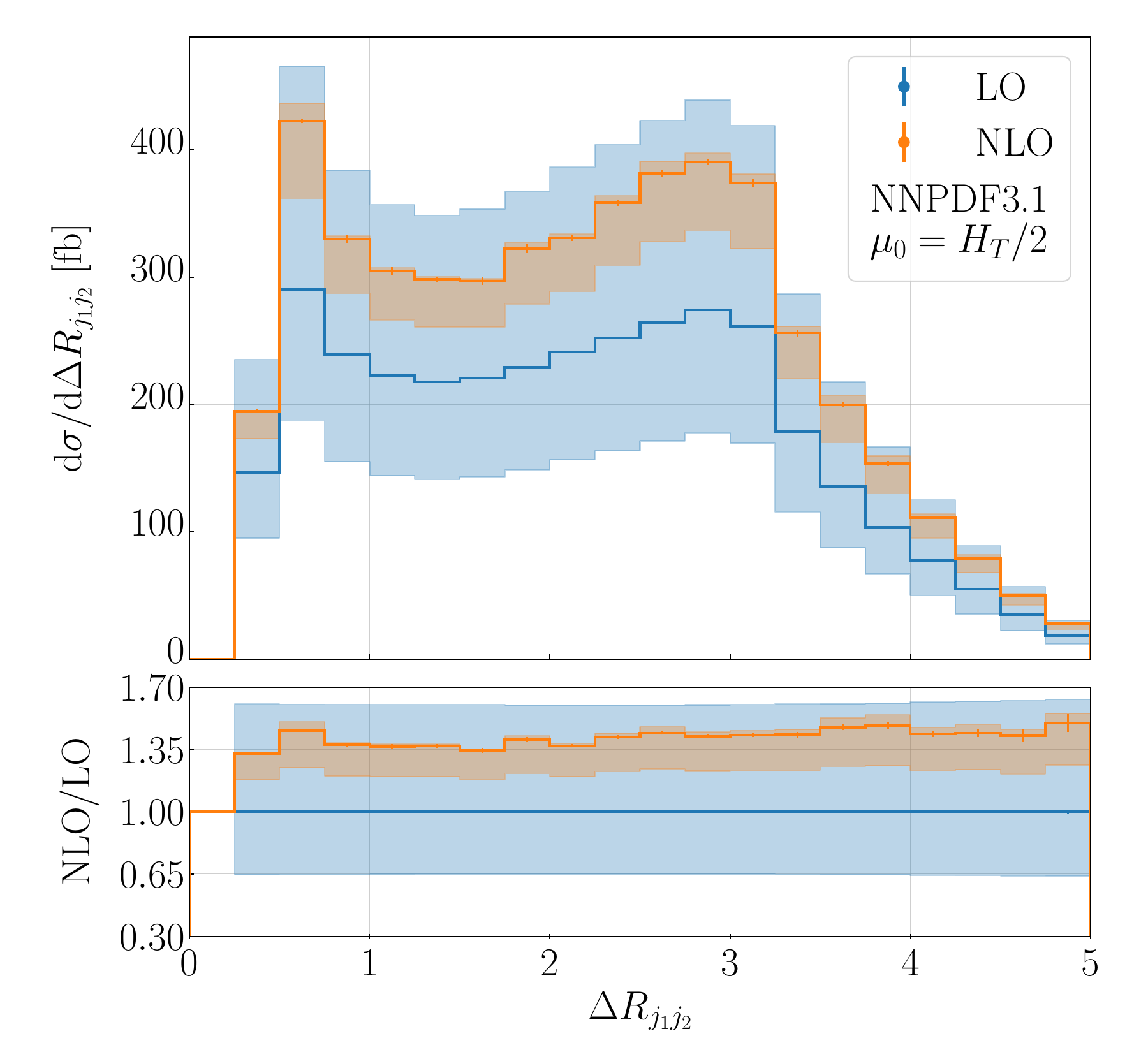}
   \includegraphics[width=0.45\textwidth]{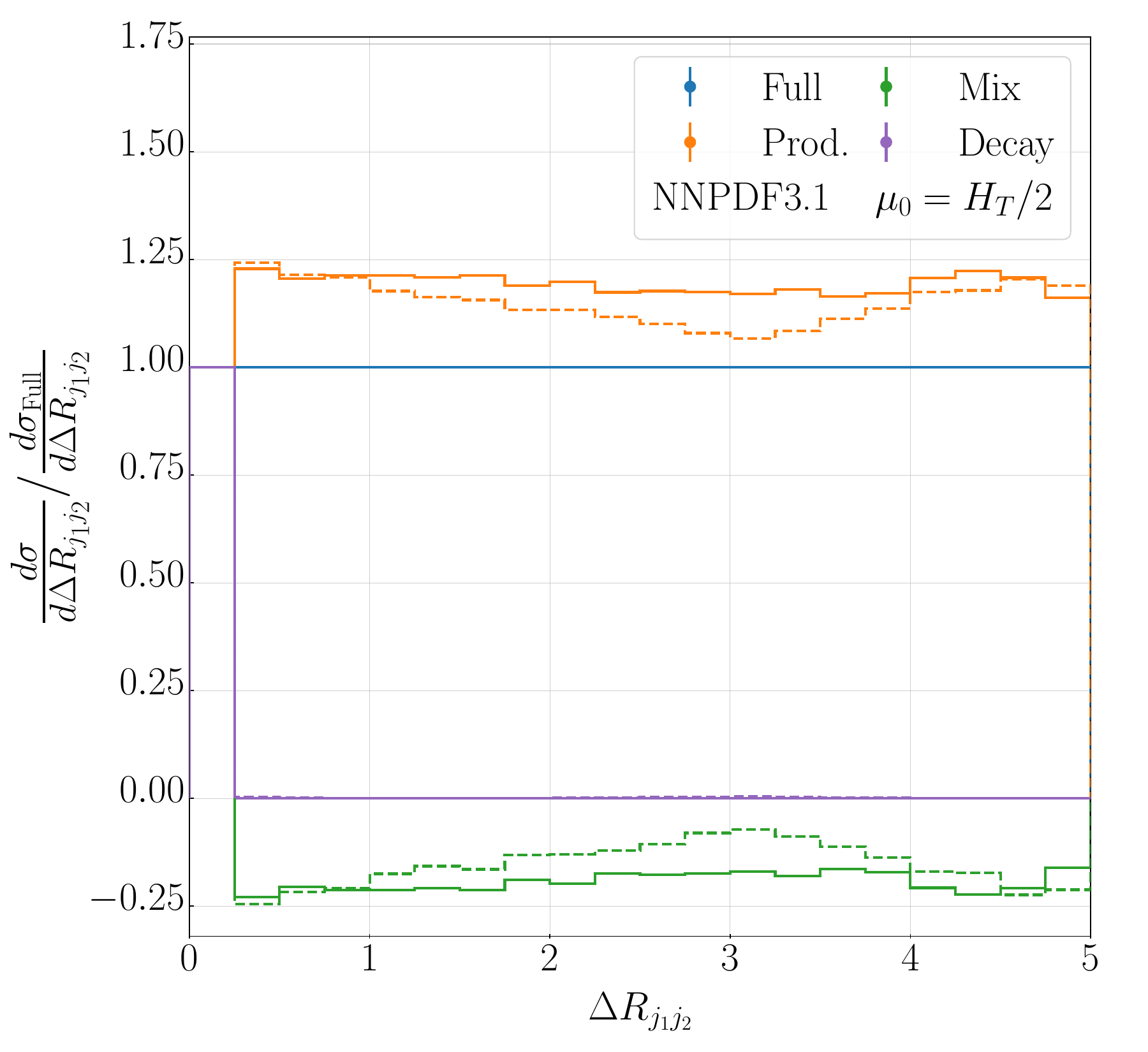}
\end{center}
\caption{\label{fig:kfac2} 
\it Same as in Figure \ref{fig:kfac1} but for the observable $\Delta R_{j_1j_2}$. Figures were taken from \cite{Bevilacqua:2022ozv}.}
\end{figure}

At last we discuss the size of NLO QCD corrections and of the different resonant contributions at the differential level for the observables $H_T^{had} =  \sum_{i=1}^2 p_{T {j_i}} + \sum_{i=1}^2 p_{T {b_i}}$ and $\Delta R_{j_1j_2}$ shown in Figure \ref{fig:kfac1} and \ref{fig:kfac2}. On the left side we display the LO (blue) and NLO QCD (orange) predictions and on the right side we show the relative size of the different resonant contributions {\it Prod.} (orange), {\it Mix} (green) and {\it Decay} (purple) for $\Delta R_{jb} >0.8$ (solid lines) and $\Delta R_{jb} >0.4$ (dashed lines). We find for $H_T^{had}$ NLO QCD corrections of about $30\%-60\%$ which are within the LO uncertainty band. The scale uncertainties are reduced from $60\%$ to $15\%$. The relative size of {\it Mix} is the same for both setups in the tail at the level of $-20\%$ while in the beginning of the spectrum this contribution obtains large shape distortions (enhanced for $\Delta R_{jb} >0.4$) and even its sign changes. Also for angular distributions we have in general NLO QCD corrections of moderate size as shown for the observable $\Delta R_{j_1j_2}$ of about $30\%-50\%$. Again the NLO prediction lies completely in the LO scale uncertainty band. The relative size of the {\it Mix} contribution is rather flat for $\Delta R_{jb} >0.8$ over the entire range at level of $20\%-25\%$ while for the inclusive setup ($\Delta R_{jb} >0.4$) larger shape distortions in the range of $7\%-25\%$ are present in the back-to-back region at $\Delta R_{j_1j_2}\approx 3$.

\section{Cross section ratios}

\begin{table}
\centering
  \begin{tabular}{llll}
  \hline
     ${\cal R}_n$     & ${\cal R}^{\rm LO}$     &
     ${\cal R}^{\rm NLO}$ 
     &   ${\cal R}^{\rm NLO}_{\rm exp}$   \\
 &&&\\
\hline
${\cal R}_1=\sigma_{t\bar{t}j}/\sigma_{t\bar{t}}$ 
& $0.3686 {}^{\,+12\%}_{\,-10\%} $ 
&  $0.3546 {}^{\, +0\%}_{\, -5\%} $
& $0.3522 {}^{\, +0\%}_{\,-3\%} $ \\
${\cal R}_2=\sigma_{t\bar{t}jj}/\sigma_{t\bar{t}j}$
& $0.2539 {}^{\, +11\%}_{\, ~-9\%}$
& $0.2660 {}^{\, +0\%}_{\, -5\%}$ 
& $0.2675 {}^{\, +0\%}_{\, -2\%}$ 
\\
\hline
  \end{tabular}
 \caption{\it \label{tab:2}  LO and (expanded) NLO cross section ratios for the $pp \to t\bar{t}+nj$ processes. Results are given for the default cuts with $\Delta R_{jb} > 0.8$. Scale uncertainties are also shown. Table was taken from \cite{Bevilacqua:2022ozv}.}
\end{table}

Finally, we present in Table \ref{tab:2} results for the cross section ratios ${\cal R}_n=\sigma_{t\bar{t}+nj}/\sigma_{t\bar{t}+(n-1)j}$ for $n=1,2$ in the dilepton decay channel with $\Delta R_{jb} > 0.8$. We employ the central scale $\mu_0=H_T/2$ given in Eq. \eqref{eq_scale} for all processes entering the cross section ratios where we restrict the summation of $p_{T {j_i}}$ to the number of light jets present in the corresponding Born process. Scale uncertainties are calculated in a correlated way by simultaneously varying the scale in the numerator and denominator. We obtain for both ratios, ${\cal R}_1$ and ${\cal R}_2$, NLO QCD corrections of $4\%-5\%$ which are in agreement within the LO scale uncertainties of $11\%-12\%$. These uncertainties are reduced at NLO to $5\%$. In the last column we present a consistent expansion in $\alpha_s$ of this ratio at NLO QCD labeled as ${\cal R}^{\rm NLO}_{\rm exp}$. By this expansion the scale dependence is reduced to $2\%-3\%$ and the central value differs by less than $1\%$ with respect to ${\cal R}^{\rm NLO}$. The internal PDF uncertainties of the NNPDF3.1 PDF set are at the level of $0.5\%$ and thus they are significantly smaller than the theoretical uncertainties obtained by scale variation.

\section{Summary}
In this proceeding we have presented the calculation of NLO QCD corrections to the $pp \to t\bar{t}jj$ process in the dilepton decay channel at the LHC. Both NLO QCD corrections as well as jet radiation have been consistently included in the production and decay of the top-quark pair. At LO the full calculation was purely dominated by the {\it Prod.} contribution such that {\it Mix} and {\it Decay} can be safely neglected at this order. However, at NLO we have found that the different resonant contributions start to mix and because of that the {\it Mix} contribution changes its sign and increases in absolute values to about $20\%$ and thus becomes non-negligible. NLO QCD corrections at the level of $40\%$ have been found, which show good agreement within the LO scale uncertainties. In addition, these theoretical uncertainties are reduced by a factor of $4$ to $14\%$ at NLO QCD and remain still dominant compared to the internal PDF uncertainties varying between $1\%-3\%$ for different PDF sets.

\section*{Acknowledgments}
This work was supported by the Deutsche Forschungsgemeinschaft (DFG) under grant \\396021762 $-$ TRR 257: {\it P3H - Particle Physics Phenomenology after the Higgs Discovery}.


\providecommand{\href}[2]{#2}\begingroup\raggedright\endgroup

\end{document}